\documentclass[conference]{IEEEtran}
\IEEEoverridecommandlockouts
\usepackage{cite}
\usepackage{url}
\usepackage{amsmath,amssymb,amsfonts}
\usepackage{algorithm}
\usepackage{algpseudocode}
\usepackage[font={sf,scriptsize},           
            labelfont={bf,color=accessblue},
            caption=false]                  
           {subfig} 
\usepackage{booktabs}
\usepackage{comment}
\usepackage{balance}

\usepackage{booktabs,multirow}

\usepackage{graphicx}
\usepackage{textcomp}
\usepackage{xcolor}
\usepackage[T1]{fontenc}
\usepackage[utf8]{inputenc}
\def\BibTeX{{\rm B\kern-.05em{\sc i\kern-.025em b}\kern-.08em
    T\kern-.1667em\lower.7ex\hbox{E}\kern-.125emX}}

    \errorcontextlines\maxdimen

\makeatletter
    \newcommand*{\algrule}[1][\algorithmicindent]{\makebox[#1][l]{\hspace*{.5em}\thealgruleextra\vrule height \thealgruleheight depth \thealgruledepth}}%
\newcommand*{\thealgruleextra}{}
\newcommand*{\thealgruleheight}{.75\baselineskip}
\newcommand*{\thealgruledepth}{.25\baselineskip}

\newcount\ALG@printindent@tempcnta
\def\ALG@printindent{%
    \ifnum \theALG@nested>0
        \ifx\ALG@text\ALG@x@notext
        \else
            \unskip
            \addvspace{-1pt}
            \ALG@printindent@tempcnta=1
            \loop
                \algrule[\csname ALG@ind@\the\ALG@printindent@tempcnta\endcsname]%
                \advance \ALG@printindent@tempcnta 1
            \ifnum \ALG@printindent@tempcnta<\numexpr\theALG@nested+1\relax
            \repeat
        \fi
    \fi
    }%
\usepackage{etoolbox}
\patchcmd{\ALG@doentity}{\noindent\hskip\ALG@tlm}{\ALG@printindent}{}{\errmessage{failed to patch}}
\makeatother

\newbox\statebox
\newcommand{\myState}[1]{%
    \setbox\statebox=\vbox{#1}%
    \edef\thealgruleheight{\dimexpr \the\ht\statebox+1pt\relax}%
    \edef\thealgruledepth{\dimexpr \the\dp\statebox+1pt\relax}%
    \ifdim\thealgruleheight<.75\baselineskip
        \def\thealgruleheight{\dimexpr .75\baselineskip+1pt\relax}%
    \fi
    \ifdim\thealgruledepth<.25\baselineskip
        \def\thealgruledepth{\dimexpr .25\baselineskip+1pt\relax}%
    \fi
    \State #1%
    \def\thealgruleheight{\dimexpr .75\baselineskip+1pt\relax}%
    \def\thealgruledepth{\dimexpr .25\baselineskip+1pt\relax}%
}
\begin{document}

%


\title{Privacy-Enhancing Technologies in Federated Learning for the Internet of Healthcare Things: \\A Survey\\
\thanks{}
}


\author{
    \IEEEauthorblockN{Fatemeh Mosaiyebzadeh\IEEEauthorrefmark{1}, Seyedamin Pouriyeh\IEEEauthorrefmark{2}, Reza M. Parizi\IEEEauthorrefmark{3}, 
    Quan Z. Sheng\IEEEauthorrefmark{4}, Meng Han \IEEEauthorrefmark{5}, 
   \\ Liang Zhao\IEEEauthorrefmark{2},
     Giovanna Sannino \IEEEauthorrefmark{6}, Daniel Macêdo Batista\IEEEauthorrefmark{1}}
     \IEEEauthorblockA{\IEEEauthorrefmark{1} Department of Computer Science, University of São Paulo, Brazil
    \\ \ \{fatemehm, batista\}@ime.usp.br}
    \IEEEauthorblockA{\IEEEauthorrefmark{2} Department of Information and Technology, Kennesaw State University, Marietta, GA, USA
    \\ \ \{spouriye, lzhao10\}@kennesaw.edu}
    \IEEEauthorblockA{\IEEEauthorrefmark{3}Decentralized Science Lab, Kennesaw State University, Marietta, GA, USA   \\ rparizi1@kennesaw.edu}
     \IEEEauthorblockA{\IEEEauthorrefmark{4} School of Computing, Macquarie University, Sydney, Australia \\ michael.sheng@mq.edu.au }
     \IEEEauthorblockA{\IEEEauthorrefmark{5} Binjiang Institute of Zhejiang University, Hangzhou, Zhejiang, China \\ mhan@zju.edu.cn}
    \IEEEauthorblockA{\IEEEauthorrefmark{6} Institute of High Performance Computing and Networking, National Research Council, Naples, Italy \\ giovanna.sannino@icar.cnr.it}
}

\maketitle

\begin{abstract}
Advancements in wearable medical devices in IoT technology are shaping the modern healthcare system. With the emergence of the Internet of Healthcare Things (IoHT), we are witnessing how efficient healthcare services are provided to patients and how healthcare professionals are effectively used AI-based models to analyze the data collected from IoHT devices for the treatment of various diseases. To avoid privacy breaches, these data must be processed and analyzed in compliance with the legal rules and regulations such as HIPAA and GDPR.
Federated learning is a machine leaning based approach that allows multiple entities to collaboratively train a ML model without sharing their data. This is particularly useful in the healthcare domain where data privacy and security are big concerns.
Even though FL addresses some privacy concerns, there is still no formal proof of privacy guarantees for IoHT data. 

Privacy Enhancing Technologies (PETs) are a set of tools and techniques that are designed to enhance the privacy and security of online communications and data sharing. PETs provide a range of features that help protect users' personal information and sensitive data from unauthorized access and tracking. This paper reviews PETs in detail and comprehensively in relation to FL in the IoHT setting and identifies several key challenges for future research.


\end{abstract}


\section{Introduction}


In recent years, we have witnessed an accelerated growth of IoT devices in various domains such as healthcare~\cite{mani2020iot}, smart transportations~\cite{cheng2016fuzzy}, smart home and building~\cite{stojkoska2017review}, and smart cities~\cite{chen2022iot}. In the healthcare domain, IoT technology has shown its capabilities and applications in collecting patients' data to enable healthcare professionals to analyze the data for better and more efficient treatment of various diseases.
These devices are designed to automatically collect, send, receive, and store data over the networks in order to proactively detect, diagnose, monitor, and treat patients both in and out of the healthcare systems. 

Internet of Healthcare Things (IoHT) is a sub-type of the Internet of Things (IoT) oriented to e-health by combining various smart devices such as smart watches,  wearable trackers, and other smart connected devices to record various health measures such as heart rate, body temperature, and blood pressure~\cite{aceto2020industry}. 
A huge amount of information collected from those variety of IoHT devices and applications is later employed in data analytics where it is empowered with Artificial Intelligence (AI) and Machine Learning (ML) models to mine such information and improve the health decision making.

Traditionally, healthcare organizations use centralized ML-based models in clouds or data centers to train the data generated by IoHT devices aiming to take reliable decisions in the healthcare domain. 
However, such models usually suffer from performance and accuracy issues due to the unavailability of sufficient data to reside centrally on the server side for training due to direct access restrictions/regulations (HIPAA and GDPR) on such data, where all may lead to biased models that cannot be trustworthy~\cite{rajotte2021reducing}~\cite{vayena2018machine}. Additionally, even with sufficient data, the training procedure in a centralized setting is time-consuming and expensive tasks make them out of interest of hospitals and research centers~\cite{joshi2022federated}.

\begin{figure*}[!htb]
    \centering
    \includegraphics[width=12cm]{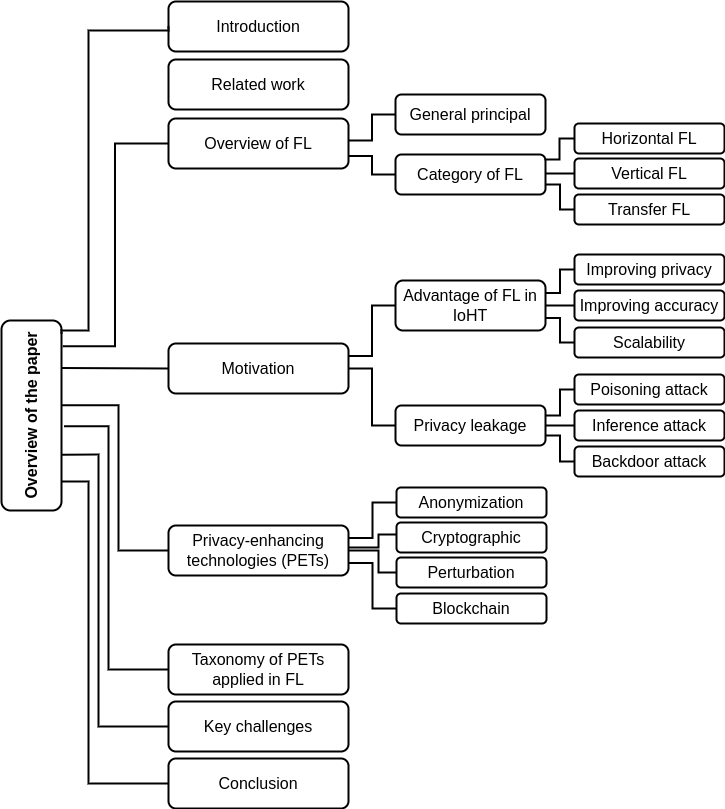}
    \caption{Outline of the paper.}
    \label{fig1}
\end{figure*}

Recently, the Federated Learning (FL) concept has been discovered as a promising way for the eHealth systems to overcome data privacy concerns in IoHT~\cite{aouedi2022handling}. FL is a distributed ML-based approach that maintains patients' data where they are generated and enables the training of ML models collaboratively on multiple clients' health data like hospitals or IoHT devices in a  decentralized network~\cite{yang2019federated}~\cite{antunes2022federated}. However, recent studies have shown that sometimes FL can not guarantee proper privacy-preserving~\cite{asad2020critical}.




Privacy Enhancing Technologies (PETs) are a set of tools and techniques that are designed to enhance the privacy and security of online communications and data sharing. PETs provide a range of features that help protect users' personal information and sensitive data from unauthorized access and tracking. 
The development of PETs can offer a reliable pathway toward data-driven technologies such as ML-based models while preserving privacy.
PETs are a group of methods, procedures, and techniques to extract value from data, and simultaneously reduce the privacy and security risk to private data~\cite{danezis2004introduction}.
PETs are crucial, especially in industries like healthcare that collect and use sensitive data extensively. In healthcare domain, collected patient data allow researchers and healthcare professionals to distinguish disease, drug development, and improve public health. For instance, vaccine development research during the COVID-19 pandemic has shown the importance of information in public health.

There are various PETs that can be utilized to improve privacy in FL. For example, secure multi-party computation (SMPC)~\cite{abou2023mitfed}, syntactic anonymization such as k-anonymity~\cite{yu2022survey}, homomorphic encryption~\cite{xue2021resource}, zero-knowledge proofs~\cite{hao2021urllc}, differential privacy~\cite{zhang2021feddpgan},  and blockchain techniques~\cite{garrido2022revealing} are some of those techniques that are aligned with FL framework and will be discussed in this paper.

In this paper, we aim to explore the privacy concerns in FL environment from PET perspective and discuss how PETs are integrated into FL to enhance privacy issues.  



To the best of our knowledge, there is no current research to provide a comprehensive survey on FL for the IoHT from a PET perspective. To fill this research gap, we comprehensively reviewed PETs and FL integration in smart healthcare environments. This paper reviews the main topics of privacy and FL in smart healthcare systems. Initially, we reviewed the privacy requirements and the cause of privacy leakage and violation in FL. Then, we review the PETs approach according to four PETs applied to FL. Finally, we summarize the PETs applied to FL and present some open issues.

The remaining part of this paper is organized as follows. Section~\ref{RELATED} summarizes the surveys similar to ours while highlighting the differences. In Section~\ref{FLH}, we provide the general principles of FL and the different sorts of this technique used in smart healthcare environments. We provide the motivations for using the privacy-preserving FL in smart healthcare in Section~\ref{PPFL}. Section~\ref{PET} is dedicated to a complete literature review of PETs. Section~\ref{PET-FL} presents PET's application on FL in the smart healthcare environment. Section~\ref{KEY} presents open issues related to PETs in FL and 
concluding remarks are given in Section~\ref{con}. Fig. \ref{fig1} 
depicts a systematic outline of this survey paper.

\section{Related Work}
\label{RELATED}
There are many review papers that cover a wide range of security and privacy challenges in FL environment that are either dedicated in other domains or covered security and privacy issues at the general level. In this section, we tried to discuss the most recent and similar to our work.

In \cite{aledhari2020related}, the focus is to 
provide an overview of FL, highlighting protocols, platforms, algorithms, market implications, and real-life use-cases, in terms of software and hardware. The advantage related to privacy, brought by FL, is presented in some parts of the work but this is not the main focus of the paper. Some use-cases related to health applications are presented but there are no comments about IoHT (In fact, authors state that IoT is not the focus of the paper).

The authors in \cite{zhang2021related} provide a formal definition of FL and review the existing works using FL. The works are evaluated in terms of five aspects and one of these aspects is privacy mechanisms. Three mechanisms are considered: {\em model aggregation}, {\em homomorphic encryption}, and {\em differential privacy}. In our survey, we focus on privacy and 
consider a different approach, classifying four techniques: {\em anonymization}, {\em cryptography}, {\em perturbation method}, and {\em blockchain}. Similar to~\cite{aledhari2020related}, in~\cite{zhang2021related} some use-cases related to health applications are presented but there are no comments about IoHT.

The authors in \cite{mothukuri2021related} review the FL method specifically in terms of both security and privacy. Different implementations of FL are considered and evaluated. Some of the FL threats in terms of security and privacy are similar to those considered in our paper. Some applications are oriented to IoT but there are no comments about IoHT.

In \cite{nguyen2021related}, the authors focused on the IoT domain only. Similar to~\cite{zhang2021related}, a formal definition of FL is presented.  Healthcare applications are considered in the survey but the comparison and analysis of the works do not specify what privacy attacks the works are oriented to and neither the datasets used by them.

In another effort~\cite{novikova2022analysis}, a number of privacy-preserving mechanisms adopted for FL frameworks are evaluated by the authors, as well as their application to vehicle activity recognition. In this study, they examined the open-source FL frameworks FATE and PFL. The FATE framework uses homomorphic encryption to secure computations and input data, while PFL uses multi-party secure computations and differential privacy to protect the processing of vertically partitioned data and train neural networks for horizontally partitioned data. Similar to~\cite{zhang2021related} and~\cite{aledhari2020related}, there
are no comments about IoHT in the survey.

Nguyen et al.~\cite{nguyen2022related} represent the summary of FL in the Internet of Medical Things (IoMT). In this study, a federated EHR management system, a federated remote monitoring system, a federated COVID-19 detection system, and a federated medical imaging system were discussed. Innovative FL designs for IoMT are investigated, including  secure FL, resource-aware FL, and incentive-aware FL. Also, privacy-enhanced FL to protect security is explored, but this is not the main focus of the paper. Similar to~\cite{novikova2022analysis}, in~\cite{nguyen2022related}, among the privacy-enhancing mechanisms, differential privacy method is taken into consideration, while in our survey, we examine four different technologies that enhance privacy.

To the best of our knowledge, this work is the first survey specifically focused on reviewing Federated Learning applications in IoHT from the perspective of Privacy-Enhancing Technologies. A side-by-side comparison of recent efforts in this domain is shown in Table~\ref{related_work}.

\begin{table*}[h!]
\centering
\small
  \caption{Comparison to the related works}
   \label{related_work}
        \begin{tabular}{lcccccccc}
    \toprule
     \textbf{References}&{\textbf{IoHT environment}}  & {\textbf{Healthcare domain}}  & \multicolumn{4}{c}{\textbf{Privacy Mechanisms}} \\
     \cline{4-7}
     \multicolumn{2}{l}{} & &\textbf{Anonymization} & \textbf{Cryptography}  & \textbf{Perturbation }& \textbf{Blockchain}\\
    \midrule
    
      \textbf{~\cite{aledhari2020related}} & $\times$ & \checkmark	& $\times$& $\times$ & $\times$&$\times$\\
    
    \midrule
    
   \textbf{~\cite{zhang2021related}} &$\times$ & $\times$  &	$\times$ & \checkmark& \checkmark& $\times$\\

    \midrule
    
     \textbf{~\cite{mothukuri2021related}} &$\times$ &$\times$  &  $\times$&\checkmark& \checkmark& \checkmark	\\

    \midrule
     
    \textbf{~\cite{nguyen2021related}} & \checkmark & \checkmark &	$\times$ 	& $\times$ &\checkmark& \checkmark \\

    \midrule
    
    \textbf{~\cite{novikova2022analysis}} &$\times$  &$\times$ &	$\times$ &\checkmark &\checkmark&$\times$ \\

    \midrule
    
    \textbf{~\cite{nguyen2022related}}&\checkmark   &\checkmark	 & $\times$ &	$\times$&\checkmark& $\times$\\
   \midrule
     \textbf{Our work} & \checkmark &\checkmark	&\checkmark  &	\checkmark  &\checkmark & \checkmark\\

\bottomrule
    \end{tabular}
\end{table*}

\section{Federated Learning for healthcare}
\label{FLH}
The overall FL principle and the many FL types in the context of e-healthcare are discussed in this section.

\subsection{Principles of FL for Smart Healthcare}
Privacy breaches 
have become a major concern for users' data. Therefore, governments establish policies to prevent privacy leakage in order to preserve users' data privacy. Breaching these policies is expensive for companies, and it has boosted the development of FL in 2016 by Google~\cite{mcmahan2017communication}~\cite{konevcny2016federated}. FL or collaborative learning 
trains a global machine learning model without explicitly exchanging the local data on multiple parties. In an FL  system, clients train local machine-learning models on local datasets and exchange some parameters like gradients or model weights with the central server to obtain a global model. 

In general, the FL process for IoHT consists of the following steps:
\subsubsection{Initialization} The aggregation server selects data generated by IoHT devices, such as Blood Sample Reader or human motion detection to do the prediction or classification task.
Furthermore, the central server chooses a group of participants to participate in the FL process.
\subsubsection{Updating  Local Training model} The server sends an initial model to the devices to initiate the distributed training after choosing the IoHT devices for the learning process. Each device computes its updated model by training a local model with its own dataset that is kept locally.
Then, each device sends its updated model to central the server in order to aggregate all of the updated models.
\subsubsection{Model Aggregation} After receiving the parameters from each IoHT device in the FL process, the aggregation step combines all parameters to generate a global learning model. Federated Averaging (FedAvg) algorithm is an averaging model which we can use to calculate the global model and send it to all IoHT devices for updating the local model.
\subsection{FL Types for Smart Healthcare}
Studying the FL methods utilized in various domains shows that based on data partitioning, FL methods can be categorized into Horizontal FL, Vertical FL, and Federated Transfer Learning.

In horizontal FL, or sample-based federated learning, the datasets of different healthcare clients have the same feature space and different sample space.
Since the local data are in the same feature space, local healthcare participants can train the local model using their local data by the same AI model such as the neural network model. Afterwards, the global model simply can be updated by combining all the local models transmitted from local healthcare organizations or institutions~\cite{mammen2021federated}. A horizontal FL example in smart healthcare can be multiple implanted medical devices from different hospitals as clients, 
that collect very similar data but have little to no overlap of patients~\cite{pfitzner2021federated}.

In vertical FL, the datasets of different healthcare organizations have similar sample spaces and different feature spaces. This method can be used to address the overlapping sample at distributed clients. The vertical FL usually utilizes entity alignment techniques to collect the overlapped samples of the hospitals. Then, the overlapped data can be applied to the local training model integrated with encryption techniques~\cite{zhang2022introduction}. An example of vertical FL in IoHT applications can be the shared learning model between hospitals and cardiologists. Both hospitals and cardiologists with various data features, which have patients with a similar sample space, use a vertical FL for training an AI model by utilizing historical medical records at hospitals and cardiologist data for smart healthcare decisions~\cite{shyu2021systematic}.

Federated Transfer Learning is an integration of transfer learning into federated learning to handle datasets that have various sample spaces and various feature spaces. In fact, transfer learning is a way to transfer knowledge from one particular problem to another
to decrease the distribution divergence between different domains~\cite{chen2020fedhealth}. An example of federated transfer learning in healthcare organizations can be disease diagnosis by collaborating countries with numerous hospitals that have various patients and various therapeutic programs~\cite{nguyen2022federated}.

\section{Privacy requirements for IoHT systems}

For IoHT devices, privacy requirements are more stringent than for typical IoT infrastructures. IoHT healthcare systems have some privacy requirements, such as data protection privacy~\cite{huang2016wearable}. 
During the collection and storage of patient data, we must continually take into account ethical privacy regulations throughout the entire data lifecycle. For instance, Privacy policies such as GDRP and HIPAA are laws for preserving privacy at the data level~\cite{mooney2018hipaa}. According to privacy policies, only authorized individuals can have access to patient health data. In fact, data privacy protection is a way to preserve personal data from unauthorized use and manipulation.

Thus, to protect the privacy of patient data, the IoHT system should be designed to guarantee the following~\cite{barrows1996privacy}:
\begin{itemize}
    \item Preserve the privacy of patients and the confidentiality of patient health care data (prevention of unauthorized access to health information).
    \item The integrity of healthcare data (prevention of unauthorized data manipulation).
    \item The availability of health data for authorized people.
\end{itemize}

\section{Motivation of using Privacy-Preserving FL in smart healthcare}
\label{PPFL}

In the next two sections, we will summarily cover the benefit FL and the potential threats of using FL in smart healthcare.

\subsection{Benefits of FL in IoHT}
Due to various characteristics of FL such as privacy-preserving, and collaborative learning in a distributed data environment  bring many advantages to the IoHT domain that will be discussed briefly in the next subsections.
\subsubsection{Improving the privacy of user data}
With increasing the number of IoHT devices and publicly available medical datasets generated by IoHT devices, privacy concerns are also growing in the e-healthcare systems. Collected data by IoHT devices, such as heartbeat, blood pressure, and glucose level, is more sensitive compared to other types of data. According to data privacy protection legislation, private patient data is the most sensitive data and is restricted by government laws. To address data privacy challenges in the e-healthcare domain, FL offers a decentralized training mechanism where each client or institution can control private data and define a privacy-preservation policy~\cite{ng2021federated}.
In the FL framework, the raw health data are stored at a medical devices or local site and do not leave the IoHT devices during the federated data training process. During model training, only the local updates like model gradients need to be sent to the central server, which reduces the risk of sensitive and personal data leakage and ensures a high level of patient data privacy~\cite{zhao2020privacy}.

\subsubsection{Less biased model}

Because a centralized model can only be trained using limited data from a single hospital, the result may be biased in the predictions. Therefore, mitigation bias recently gains a lot of attention in modern machine learning techniques~\cite{wahab2021federated}. Thus, for models to be more generalizable, more data must be used, which can be achieved through data sharing between organizations. However, exchanging patients' electronic health data between hospitals is against their data security and privacy because healthcare data is sensitive~\cite{sheller2020federated}. Under these circumstances, to address the bias issue, federated learning has emerged as an option for building a collaborative learning model for healthcare data and producing models that yield unbiased results. The trained model is less biased and smarter as different datasets from various sources are integrated into the learning process~\cite{rieke2020future}.



\subsubsection{Improving the scalability}
In the centralized model, uploading all of the healthcare data to the server leads to a waste of computing resources, breaches privacy, and puts more pressure on the wireless communication network, which declines the network's scalability.
However, FL's distributed nature enables the scalability of IoHT networks~\cite{lim2020federated}. In fact, FL has the ability to use the computational resources located at multiple IoHT devices across different hospitals located in different geographic regions in a parallel manner. For instance, when new hospitals or healthcare institutions participate, they add more computational resources in the federated learning process. Therefore, these more computational resources allow federated learning to enhance performance. Moreover, the FL architecture avoids sending the massive amounts of IoHT data gathered to the cloud, which can result in significant network bandwidth saving and drastically reduce communication costs~\cite{khan2021federated}~\cite{khan2020federated}.
 
\subsection{Privacy Leakage and Potential Threats in Federated Learning}

Although FL provides a privacy-aware framework to train a global model without sharing data and allows clients to use the framework using their local dataset, 
recent works have shown that FL can face privacy breaches and information leakage.

The FL frameworks restrict sharing data on local devices with third-party or central servers. Nonetheless, it is possible to reveal sensitive information through the back-tracing of gradients and update the communication models through the training process~\cite{melis2019exploiting}\cite{bhowmick2018protection}. 
For example, Zhu et al.~\cite{zhu2019deep} introduces a deep leakage from Gradient (DLG), which shows malicious attackers can steal the training data in a few iterations. In their study, they showed how private training data can be easily leaked because of sharing the gradients. Similarly, Aono et al.~\cite{aono2017privacy} reported that accessing a small portion of original gradients may cause leakage in local training data. 
Although FL models on decentralized data sources have shown promising results with respect to preserving data privacy. But, it is still vulnerable to several types of attacks
such as poisoning attacks~\cite{bhagoji2019analyzing}, inference attacks~\cite{ying2020privacy}, and backdoor attacks~\cite{bagdasaryan2020backdoor}.

\begin{figure}[!tb]
    \centering
    \includegraphics[scale=0.45]{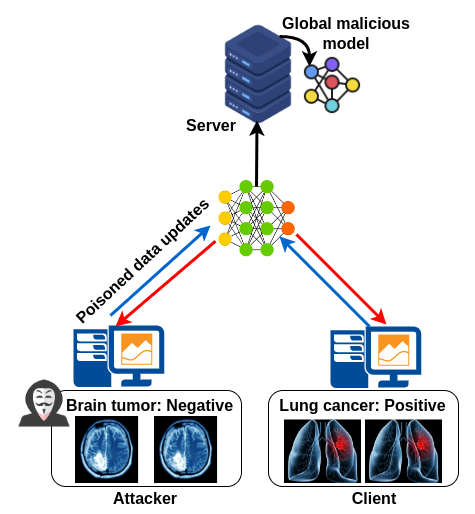}
    \caption{An illustration of the poisoning attacks against FL}
    \label{fig:poison-att}
    \centering
\end{figure}

In a poisoning attack, which occurs during the training time, an attacker tries to manipulate the training data sample by injecting designed samples to compromise the whole learning process~\cite{zhou2021deep}.
In Poisoning attacks including data poisoning attack~\cite{tolpegin2020data} and model poisoning attack~\cite{cao2019understanding} the ultimate goal of attackers is to change the behavior of the target model. A data poisoning attack aims to mislead the global model by manipulating the local training data. The attacker flips the labels of training data and adds noise in order to degrade the quality of models~\cite{lyu2020threats}. Fig.~\ref{fig:poison-att} shows how an attacker changes the trained model by flipping the data labels.
In the model poisoning attack, the attacker attempts to manipulate local model updates before sending the models to the server. This method includes various techniques to manipulate the FL local training procedure, such as direct gradient manipulation and changing the learning rule~\cite{jere2020taxonomy}.

\begin{figure}[!tb]
    \centering
    \includegraphics[scale=0.45]{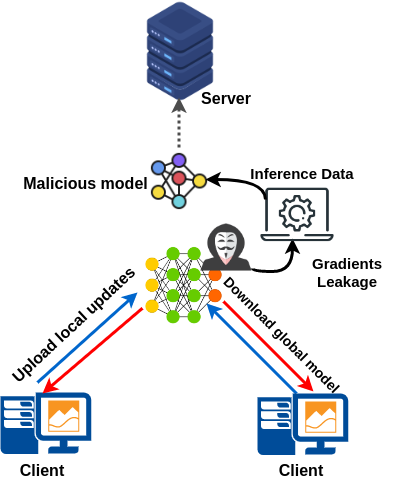}
    \caption{An illustration of the Inference attacks against FL}
    \label{fig:infer-att}
    \centering
\end{figure}

In 
an 
inference attack, the attacker aims to exchange gradients during the FL training process, which can result in serious information leakage about the features of clients’ training data. The inference attack includes inferring class representatives~\cite{sun2021soteria}, inferring membership~\cite{zhang2020gan}, inferring data properties~\cite{wang2022poisoning}, and inferring samples/labels~\cite{duan2022combined}. 
In the inference of class representatives, the adversary creates samples that are not in the original training dataset. Attackers use these false samples to learn sensitive information about the training dataset~\cite{hitaj2017deep}. The inference of memberships tries to determine whether a given data sample
has 
been used for model training~\cite{wang2019beyond}. 
In the property inference attack, the attacker aims to infer the property information of the training dataset~\cite{shen2020exploiting}. In the inferring samples, the attacker recreates labels from the gradients and recovers the original training samples that 
are 
used during training~\cite{wei2020framework}. Fig~\ref{fig:infer-att} shows an example of inference attacks.

\begin{figure}[!tb]
    \centering
    \includegraphics[scale=0.45]{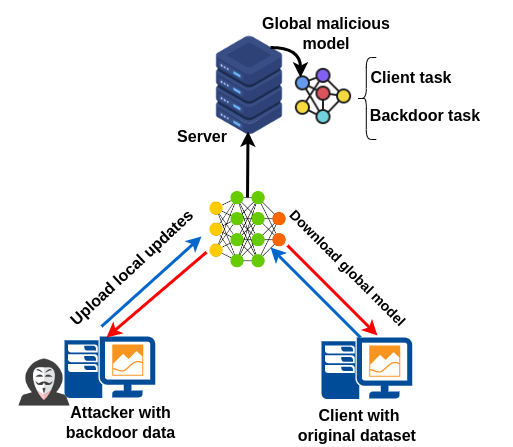}
    \caption{An illustration of the Backdoor attacks against FL}
    \label{fig:back-att}
    \centering
\end{figure}

In  a backdoor attack, the goal of the attacker is to destroy the global FL model and then replace the actual global FL model with the attacker's model~\cite{zeng2022never}. This attack can be also classified as a type of model poisoning attack, but it is 
more harmful than poisoning attacks~\cite{bagdasaryan2020backdoor}. In fact, the attacker compromises the devices of one or several participants and trains a model using
backdoor data, then submits the resulting model. After federated averaging, the global model is replaced by the backdoored model as shown in Fig.~\ref{fig:back-att}.  
In the backdoor, an adversary can be hidden and has no impact on the accuracy or functionality of the global model like accuracy. As a result, the accuracy of the validation dataset makes it difficult to distinguish the backdoor attack~\cite{yin2021backdoor}\cite{sun2019can}.

\section{Privacy enhancing technologies}
\label{PET}
Privacy Enhancing Technologies (PETs) are a set of tools and techniques that aim to protect individuals' privacy. PETs are designed to enable companies to embed privacy-by-design principles into their data governance practices to minimize the amount of personal data they collect, use and share while maximizing data security and privacy. In this context, our objective is to explore how PETs can be utilized to enhance privacy-preserving in FL to improve patient data privacy in IoHT devices and e-healthcare.
Four wide categories of PETs are used to
improve privacy protection, including 
(1) anonymization technique~\cite{fischer2001security}, 
(2) cryptographic technique~\cite{abbas2014review}, 
(3) perturbation technique~\cite{parra2014privacy}, and  
(4) blockchain technique~\cite{liu2020enhancing}.

\subsection{Anonymization Techniques}
Anonymization techniques are broadly used for privacy enhancing by changing the 
state of a data set and removing the identifier from dataset information in a way so
that the dataset is usable and protects the privacy of individual's personal
information~\cite{dhiman2022federated}.
Anonymity technology can better avoid the leakage of sensitive patient data and
provide more secure environment for smart healthcare systems. There are several anonymization technologies that are appropriate for big medical data, which are based on three categories of widely used
anonymity protection techniques: {\em k-anonymity}, {\em l-diversity}, and {\em t-closeness} models~\cite{samarati1998protecting}.

The idea of k-anonymity is to anonymize the quasi-identifier
in the dataset that can be used by attackers to identify
sensitive information about individuals. After selecting the
quasi-identifiers, k-anonymity applies for each sample in
the dataset, which can guarantee that each sample in the
dataset cannot be re-identified from at least $k$-1
samples~\cite{choudhury2020anonymizing}. l-diversity is an
extension of the k-anonymity mechanism to enhance privacy
against against homogeneity attacks and background knowledge attacks on k-anonymity~\cite{machanavajjhala2007diversity}.
l-diversity ensures that there are at least $l$
“well-performing” values for the sensitive attributes and
protects against attribute
disclosure~\cite{sei2017anonymization}. 
Finally, t-closeness is
proposed to reduce attacks against
k-anonymity and l-diversity approaches and solve the
attribute disclosure problem~\cite{domingo2015t}.

\subsection{Cryptographic Techniques}
Cryptographic techniques 
have been used to avoid privacy disclosure of individual's private data in federated learning~\cite{blanco2021achieving}. 
These methods consist of {\em homomorphic encryption}, {\em secure multi-party computation}, and {\em zero-knowledge proofs}.

Homomorphic encryption is a form of encryption for privacy-enhancing in FL to prevent information leakage during the parameter exchanging process among clients. In this method, parameters are encoded before adding or multiplying operations~\cite{li2020review}. There are two main widely used homomorphic encryption methods: fully homomorphic encryption and partially homomorphic encryption. Fully homomorphic encryption supports both additive and multiplicative operations on ciphertext, while partially homomorphic encryption only supports either additive or multiplicative operations on the ciphertext. Compared with partially homomorphic encryption, fully homomorphic encryption provides stronger encryption, and both can be applied to horizontal and vertical federated learning~\cite{liu2022distributed}.

Secure multi-party computation (SMC)~\cite{yao1982protocols} is a sub-field of cryptographic schemes to protect private information. SMC can be used to solve the problem of collaborative computing between all parties such that no party learns anything about other participants' data~\cite{yin2021comprehensive}. 
The application of SMC  
allows multiple participants 
to concentrate on safely calculating a function for various participant 
without the requirement of trusted third-parties and revealing input. However, due to the additional encryption and decryption operations, SMC suffers from computational overhead and high communication costs~\cite{ma2022applying}.

Zero-knowledge proofs (ZKP)~\cite{goldwasser1989knowledge} is a cryptographic system to achieve both input privacy and verifiability in federated learning~\cite{guo2020v}. A zero-knowledge proof involves 
a 
prover to make sure with another entity called a 
verifier
that distinguishes the validity of a given statement. ZKP can be an appropriate method for verification of sensitive healthcare data among collaborators because it allows sharing 
data securely and privately between multiple participants~\cite{liu2019secure}.

\subsection{Perturbation Techniques}
A perturbation method is to protect private data and model privacy by adding random noise to the original data or training data during the training process. The differential privacy technique is a widely used perturbation method implemented in the FL frameworks in medical applications. It is one of the PETs methods and guarantees privacy~\cite{li2020privacy} using probability statistical models to mask sensitive private data in a dataset~\cite{jordan2022selecting} and protect healthcare data against inference attack on FL frameworks. By adding noise to the model parameters or data, data can be deferentially private~\cite{xiao2010differential}\cite{song2013stochastic}, and the parties cannot
realize whether an individual record participates in the learning process or not.

Differential privacy techniques include two categories: {\em global differential} and
{\em local differential} privacy techniques. In the global differential privacy (GDP)
setting, there is a trusted curator 
that applies carefully random noise to the real
values returned for a particular query~\cite{chan2012differentially}. Different
from GDP, a local differential privacy (LDP) technique does not need a trusted
third-party. In fact, LDP allows users to locally perturb the input data, and it
often produces too noisy data, as noise is applied to achieve individual record
privacy~\cite{cormode2018privacy}. As an advantage, the differential privacy technique by adding random noise makes data sets more secure because an attacker cannot distinguish which information is true. Therefore, more noises that are added to the sensitive data have a direct relationship to how the data is hard for an attacker to recognize true information about individuals in the dataset~\cite{kaaniche2020privacy}.

\begin{table*}[!tb]

\centering

\caption{Summary of anonymization techniques applied in FL for the smart healthcare environment.}
\label{anonymity}
\begin{tabular}{ccccccc}
\hline
\textbf{Ref.} & \textbf{Aim} & \textbf{Dataset} & \textbf{Dataset Available} & \textbf{Open-Source}&\textbf{Privacy Attack}& \textbf{{Privacy-Enhancing Method}}\\

\hline
\cite{choudhury2020syntactic} &Maximize data utility& MIMIC III &\checkmark&$\times$&Inference attack  & syntactic anonymization
\\
&\&& &&&&\\
&model performance& &&&&\\
\hline
 \cite{grama2020robust} &Applying data privacy& Pima Indians&\checkmark&$\times$
 &Poisoning attack& k-anonymity\\
 &engineering without&diabetes&&
 && \\
 &reducing the accuracy&\&&&&& \\
 &&Cleveland &&&
 & \\
  &&heart disease&&
 && \\

\hline
 \cite{alsulaimawi2020non}&Preserving private data   &MNIST&\checkmark&$\times$& - & Non-negative matrix\\
&with high accuracy& HARUS &&&&factorization\\
\hline
 \cite{cui2021fearh}&Avoiding an attack from& eICU &\checkmark&$\times$&Inference attack&Anonymous random\\
&an untrustable central &&&&&hybridization  \\
&analyzer in FL &  &&
\\
&\&  &  &&
\\
& obtain similar performance&&
\\
& compared with a centralized &&
\\
& model&&\\
\hline

\end{tabular}
\end{table*}

\subsection{Blockchain Techniques}

Blockchain is beneficial in many non-financial industries such as healthcare due to its cryptographic security, immutability, and accountability~\cite{javed2021petchain}.
Researchers have recently started implementing blockchain technology to decentralize traditional data management systems. For instance, blockchain-based data management prevents security breaches and assure GDPR compliance~\cite{truong2019gdpr}. Therefore, blockchain-based PETs solutions can be used in Medical IoT to safeguard individuals’ rights over their personal data~\cite{alamri2020preserving}. Accordingly,
Blockchain is a promising technique to improve the security and scalability of the FL system. This technique has provided a high level of security in the domain of healthcare by integrating blockchain into a federated learning to maintain the trained parameters~\cite{lu2020blockchain}. The blockchain-based system is effective for decentralized federated learning training without the need for any central server which can mitigate risks of single-point failures~\cite{nguyen2021federated_blockchain}. To provide IoHT data provenance, blockchain has shown great promise, and 
also provides permission control of the participants
to enhance the security and privacy of
parameters in federated learning. 
Blockchain has gained popularity for managing the trust and provenance of trustworthy federated nodes, their datasets, the accuracy of the models, and the immutability of the global model~\cite{hu2020survey}. A blockchain method consists of public (permissionless), private and consortium (permissioned). A public blockchain system allows any client to participate in the decentralized process without the need for authorized permission. In a private and consortium system, only the client with authorized permission can be involved in the block validation and confirmation process.

\section{PETs in Federated Learning}
\label{PET-FL}
In this section, we discuss the security and privacy issues in FL from PETs perspective. The PETs used in FL can be 
classified in several categories detailed as follows. 

\subsection{Anonymization Methods}
Several research has been published in the literature that integrates anonymization techniques and FL~\cite{orekondy2018gradient}~\cite{hao2022waffle}~\cite{song2020analyzing}, and some of these studies attempt to evaluate the incorporation of FL and anonymization methods in a smart healthcare environment~\cite{marulli2021evaluating}. For instance, in~\cite{choudhury2020syntactic}, the authors proposed a syntactic anonymity approach to guarantee privacy in federated learning. They used the anonymization based on ($k$, $k^n$)-anonymization algorithm. This approach contains two steps. In the first step, the anonymization method is applied to the original private data, which includes relational and a transactional attributes, at the local site and then feeds this anonymized data to a global model.  The second step is a global anonymization mapping process, which can be used for the prediction process in the FL global model.
They evaluated the proposed method using  Medical Information Mart for Intensive Care (MIMIC III)\footnote{https://registry.opendata.aws/mimiciii/} dataset gathered from 
one 
million patients. The results demonstrated that this approach enhances the level of privacy compared to the differential privacy method in FL.

\begin{table*}[!t]

\centering

\caption{Summary of Cryptographic algorithms applied in FL for the smart healthcare environment}
\label{crypto}
\begin{tabular}{ccccccc}
\hline
\textbf{Ref.} & \textbf{Aim} & \textbf{Dataset} & \textbf{Dataset Available} & \textbf{Open-Source}& \textbf{Privacy Attack}& \textbf{{Privacy-Enhancing Method}}\\

\hline
\cite{zhang2022homomorphic} &Preserving privacy in   & HAM10000&\checkmark&$\times$&Inference attack & Homomorphic encryption \\
&skin cancer detection& &&&&  \& \\
&with reliable accuracy& &&&&secure multi-party computation\\

\hline
 \cite{ma2022privacy} &Securing FL environ- & UP‐FALL&\checkmark&$\times$
 &Inference attack&xMK‐CKKS multi‐key \\
&ment on IoHT devices &&&&& homomorphic encryption\\
\hline
\cite{stripelis2021secure}&Enhancing patient's   &3D brain MRI&\checkmark&$\times$& Membership  & fully homomorphic  \\
&data privacy&  &&&inference attack&encryption (FHE)\\
\hline
\cite{rachakonda2022privacy}&Protect medical data   & eICU &\checkmark&$\times$&Reverse &Secure multi-party computation\\
&privacy on IoHT devices &  & &&engineering& \\
& &  & &&attack&\\
\hline
\cite{heiss2022advancing}&Privacy-enhanced  & Daily and Sports &\checkmark& $\times$&Global aggregation&Zero-knowledge proofs\\
& decentralized & Activities& &&\&&\\
&applications &  &&&poisoning attack&  \\
\hline
\end{tabular}
\end{table*}

Similarly, Grama et al.~\cite{grama2020robust} presented an adaptive privacy-preserving FL method for healthcare data. In order to enhance privacy, they used the k-anonymity method on top of the FL that can protect data by anonymization. In fact, anonymization by applying the data protection method can cause 
information loss. 
But, the proposed k-anonymity method in this paper decreases losing data. Similar to~\cite{choudhury2020syntactic}, the authors evaluated the performance of the proposed approach based on two health datasets to predict diabetes mellitus onset\footnote{https://www.kaggle.com/datasets/uciml/pima-indians-diabetes-database} and heart failure diseases\footnote{https://archive.ics.uci.edu/ml/datasets/heart+disease}. Their results showed that the k-anonymity method using $k$=4 can improve the protection of the healthcare data, if it is applied to a dataset that contains an adequate number of samples.

In~\cite{alsulaimawi2020non}, the author presented a federated PF-NMF framework. This FL framework contains multiple local privacy filters (PF), which are used to remove sensitive data to minimize the risk of privacy leakage. In the training phase, PF acts as an encoder. The framework includes a decoder in the testing phase and feeds the test data into the autoencoder. 
The author 
evaluated the proposed approach on the MNIST \footnote{https://www.tensorflow.org/datasets/catalog/mnist} and HARUS dataset (human static and dynamic activities gathered by wearable devices)\footnote{http://archive.ics.uci.edu/ml/datasets/Human+Activity+Recognition

+Using+Smartphones}. The results showed that federated PF-NMF achieves better accuracy and enhances the privacy of sensitive data.

In another work\cite{cui2021fearh}, the authors proposed a new method called federated machine learning with anonymous random hybridization (FeARH) to eliminate the privacy problems in an untrustworthy central analyzer. The hybridization algorithm adds the randomization into the parameter sets shared with other parties. With a hybrid algorithm, the medical data, which is replaced by a randomized parameters, 
do not need to be shared with other institutions. They evaluated the proposed approach on eICU dataset\footnote{https://eicu-crd.mit.edu/} and the results showed that FeARH achieves similar performance compared with FL and centralizes the machine learning method. Similar to~\cite{alsulaimawi2020non}, in \cite{cui2021fearh} the authors use anonymized data in the training phase.
Table~\ref{anonymity} 
summarizes the representative existing anonymization techniques applied for FL in smart healthcare.

\subsection{Cryptographic Algorithms}

Cryptographic methods are widely used in several FL methods to preserve data privacy when exchanging intermediate parameters during the FL training process~\cite{wibawa2022homomorphic}~\cite{bai2021method}~\cite{wibawa2022bfv}. 
For example, similar to \cite{cui2021fearh} which covers smart healthcare domain,
Zhang et al.~\cite{zhang2022homomorphic} presented the FL mechanism in the Internet of Healthcare Things (IoHT). They applied the cryptographic masking scheme based on homomorphic encryption and the secure multi-party computation to protect private medical data against reconstruction attacks or model inversion attacks. To evaluate the efficiency of the proposed FL model and validity of the privacy-enhancing masking scheme, the authors used real skin cancer datasets\footnote{https://dataverse.harvard.edu/dataset.xhtml?persistentId=doi:10.7910/
DVN/DBW86T}. The result showed that the proposed model improves the privacy protection of the medical data and achieves reliable accuracy in skin cancer detection.

Ma et al.~\cite{ma2022privacy} proposed a novel privacy-enhancing FL on a smart healthcare scenario for elderly‐fall detection, the authors used UP-FALL Detection dataset\footnote{http://sites.google.com/up.edu.mx/har-up/}. Similar to~\cite{zhang2022homomorphic}, they applied homomorphic encryption scheme to FL in order to prevent privacy leakage and achieve secure encryption
and decryption in the FL system. The proposed xMK‐CKKS multi‐key homomorphic encryption scheme utilizes an aggregated public key to encrypt the model updates before sharing them with a server for aggregation. The model decryption occurs after clients share information of their secret keys. The result showed that the proposed FL scheme using multi‐key homomorphic encryption is effective in communication, computational cost, and energy consumption, 
while ensuring the implementation of secure FL on IoHT devices.

In~\cite{stripelis2021secure}, the authors combined FL and fully homomorphic encryption (FHE) to define a novel secure FL framework for biomedical data analysis. They used the CKKS  homomorphic encryption scheme based on ciphertext packing and rescaling, similarly to the authors in~\cite{ma2022privacy}. The authors evaluated the performance of the proposed FL model using a large-scale 3D brain MRI dataset\footnote{https://dataverse.harvard.edu/dataset.xhtml?persistentId=doi:10.7910/

DVN/2RKAQP} to predict brain age in a secure environment. The result showed that the integration of a FL framework and encryption scheme does not reduce the efficiency of FL, also increases the privacy of the patient's private data.

In~\cite{rachakonda2022privacy}, the authors provided a secure and scalable FL framework to implement AI across hospital sites, collaborators, and edge devices. Similarly to~\cite{zhang2022homomorphic}, to address privacy challenges, they integrated the proposed FL framework with a secure multi-party computation algorithm to avoid data leakage and reverse engineering attacks via model updates. They evaluated the performance of the SMPC method in FL using the Philips ICU dataset.  The results demonstrated that the developed FL framework with a SMPC algorithm can be used in a large ecosystem of the Internet of Healthcare Things (IoHT) and healthcare hospital sites. Moreover, the proposed framework significantly protects medical data privacy.

Heiss et al.~\cite{heiss2022advancing} proposed a model for blockchain-based FL that leverages verifiable off-chain computations (VOC) using zero-knowledge proofs (ZKP). The architecture enables the computational correctness of 
local learning processes verifiable on 
blockchain and provides globally verifiable management of global learning parameters. They evaluated the performance of the architecture through an in-home health monitoring system where sensitive data serve as inputs to the FL system. The author used Daily and Sports Activities dataset\footnote{https://archive.ics.uci.edu/ml/datasets/daily+and+sports+activities} and the results showed that verifiable off-chain computations (VOC) using zero-knowledge proofs (ZKP) enhances privacy in decentralized applications. Similar to~\cite{rachakonda2022privacy} and~\cite{zhang2022homomorphic}, in~\cite{heiss2022advancing}, the authors integrated zero-knowledge proofs (ZKP) with FL in order to enhance privacy in IoHT ecosystem. Table~\ref{crypto} 
summarizes the cryptographic methods applied for FL in smart healthcare.

\subsection{Perturbation Methods}
Similar to~\cite{ma2022privacy}, in~\cite{kerkouche2021privacy}, the authors proposed a bandwidth-efficient FL framework in IoHT environment. The framework ensures privacy  for FL based on Differential Privacy (DP). They
discovered 
that exchanging the model update from a huge amount of IoHT devices needs a significant bandwidth. 
Therefore, they proposed the FL-SIGN-DP scheme to reduce communication costs and enhance privacy. Participants in FL-SIGN-DP only transmit the updated model's sign to the aggregation server. They used the electronic health records of roughly a million patients to assess the performance of the proposed  scheme with regard to the in-hospital mortality rate. The proposed scheme is compared with centralized learning, FL-SIGN without using standard FL, differential privacy, and differential privacy with standard FL. The results showed that the FL-SIGN-DP consumes less bandwidth and can guarantee privacy protection. 

Islam et al.~\cite{islam2022privacy} proposed 
a 
FL model to analyze patients' genomic data and identify the risk of heart failure. To enhance the privacy-preserving of the patient private data sharing among collaborating healthcare organizations in FL framework, they applied differential privacy mechanisms through feature selection based on statistical methods to increase scalability and accuracy in a federated setting where data are vertically partitioned. They evaluated the performance of the proposed FL framework using the IQVIA dataset and BC-TCGA dataset\footnote{https://www.kaggle.com/datasets/saurabhshahane/gene-expression-profiles-of-breast-cancer}for predicting the causes of certain
heart failure and the BC-TCGA dataset for cancer prediction to compare their proposed FL method. The result demonstrated that their proposed model obtains better accuracy with the highest privacy for the IQVIA and BC-TCGA datasets in a federated training setting.

\begin{table*}[!t]
\centering

\caption{Summary of Perturbation method applied in FL for the smart healthcare environment}
\label{Perturbation}
\begin{tabular}{ccccccc}
\hline
\textbf{Ref.} & \textbf{Aim} & \textbf{Dataset}& \textbf{Dataset Available} & \textbf{Open-Source}&  \textbf{Privacy Attack}& \textbf{{Privacy-Enhancing Method}}\\

\hline
\cite{kerkouche2021privacy} &Enhancing privacy   & Two real-world &$\times$&\checkmark&Inference attack  & Differential privacy \\
&\&& Electronic &&\\
&bandwidth efficiency& Health Records&&\\
\hline
 \cite{islam2022privacy} &Preserving privacy and predicting  & BC-TCGA
 &\checkmark&$\times$&- & Differential privacy \\
&risk of heart failure &&&\\
\hline
\cite{zhao2021federated}&Avoiding medical data leakage   &Dataset of a & $\times$&$\times$&Adversarial
attack  &  Differential privacy  \\
&during data exchange& tumor hospitals &\\
\hline
 \cite{li2021federated}&privacy-preserving IoHT& ADReSS  &\checkmark&$\times$&Man-in-the-middle&Differential privacy  \\
&Alzheimer’s disease detection &  &&& attack&
\\
\hline
\cite{nguyen2021federated_covid}&Preserving privacy and improving   & DarkCOVID  &\checkmark&$\times$&-&Differential privacy  \\
&COVID19 detection & ChestCOVID &&
\\
\hline
\end{tabular}
\end{table*}

The authors in~\cite{zhao2021federated} proposed federated adversarial learning (FAL) on biomedical named entity recognition (BioNER). The differential privacy technology is also used to protect the security and privacy of the data, which adds Gaussian noise during the local training and model aggregation process to enhance privacy.  More specifically, only the noised parameters with differential privacy are transferred among the server and the client. Therefore, the data leakage possibility has decreased on the local client's side. The dataset collected from 5 departments of a tumor hospital is used to examine the performance of the proposed scheme. The Result showed that the proposed FAL framework can connect data parties and prevent data leakage during data exchange inside medical institutions.

Similarly to~\cite{kerkouche2021privacy}, in~\cite{li2021federated}, the authors proposed a cost-effective
and privacy-preserving FL framework which is  IoHT Alzheimer’s disease detection scheme.
They presented an FL based privacy-preserving
smart healthcare system, namely ADDetector, to detect Alzheimer’s
disease. Moreover, they implemented a differential privacy (DP)
mechanism on the user data to avoid patient's data leakage during
transferring data to the client and enhance the privacy level
against the attacker. An ADReSS Challenge dataset from INTERSPEECH
2020\footnote{https://luzs.gitlab.io/adress/} is used to evaluate the performance of the ADDetector FL-based
scheme. The proposed FL-based framework and DP-based mechanism use
the audio from smart devices to detect low-cost Alzheimer’s
disease. The experimental results showed that the ADDETECTOR
FL-based framework achieves better accuracy and low average time
overhead with a high level of privacy and security protection.

Dinh et al.~\cite{nguyen2021federated_covid} proposed an FL framework, called FedGAN, to facilitate COVID-19 detection by enhancing privacy among medical institutions in edge cloud computing. The aim of the framework is to create realistic COVID-19 X-ray data and detect it automatically without the need for sharing COVID-19 image with parties. Additionally, they integrated a differential privacy at each hospital site to increase and guarantee the data privacy in federated COVID-19 data training. To apply the differential privacy, they used both differentially private stochastic gradient descent and a gradient perturbation technique; they also added the Gaussian noises to the gradient during the training. Additionally, they use the FedGAN blockchain-based system for safe COVID-19 data analysis. To evaluate the performance of proposed FedGAN model, they used two popular COVID-19 X-ray data sets for simulations, including a DarkCOVID\footnote{https://github.com/ieee8023/COVID-chestxray-dataset} and a ChestCOVID\footnote{https://github.com/ieee8023/covid-chestxray-dataset} dataset. The result demonstrated that FedGAN framework enhances the performance of COVID-19 detection and provides high level of privacy. Table~\ref{Perturbation} presents a summary of the perturbation methods applied for FL in smart healthcare.

\begin{table*}[!tb]
\centering

\caption{Summary of Blockchain method applied in FL for the smart healthcare environment}
\label{blockchain}
\begin{tabular}{ccccccc}
\hline
\textbf{Ref.} & \textbf{Aim} & \textbf{Dataset} &\textbf{Dataset Available} & \textbf{Open-Source}& \textbf{Privacy Attack}& \textbf{{Privacy-Enhancing Method}}\\

\hline
\cite{samuel2022iomt} & IoMT privacy-preserving &- &$\times$&$\times$&Backdoors&Blockchain  \\
&\&&&&&\&\\
&predict the COVID-19&&&&Inference attack\\
\hline
 \cite{aich2022protecting} &Preserving data privacy& - &$\times$&$\times$& -& Blockchain \\
 &\&&&&\\
&predicting COVID-19&&&\\
\hline
\cite{lakhan2022federated}&  Fraud-detection for  &ECG heartbeat &$\times$&$\times$&Fraud attack&  Blockchain\\
&IoHT&  E-Heart videos &\\
&&  Blood pressure &\\

\hline
 \cite{singh2022framework}&Privacy preserving for& Healthcare data &$\times$&$\times$&Replay attack& Blockchain \\
&IoHT in cloud &  & &
\\
\hline
\cite{kumar2021blockchain}&Preserving the patient's & CC-19  &\checkmark&\checkmark&-& Blockchain \\
& privacy and detection &  &&\\
&  COVID-19 CT scan &  &&\\
\hline
\end{tabular}
\end{table*}

\subsection{Blockchain Methods}
\label{subsec:blockchain}
Blockchain methods have been widely used in many FL methods to provide privacy and security in IoHT (smart healthcare systems).

 For smart healthcare systems, the authors of~\cite{samuel2022iomt} proposed an infrastructure called FedMedChain, which is based on secure FL and blockchain to predict the COVID-19 for IoMT scenarios, similarly to~\cite{nguyen2021federated_covid} which proposes an privacy-preserving FL-based scheme for analysis of COVID-19 data in a secure environment. The proposed system can improve public communication and address the challenges of big data silos and data security. Furthermore, information security and privacy analyses showed that the proposed infrastructure is robust against privacy breaches and can improve information security.

Similarly to~\cite{samuel2022iomt}, to solve privacy concerns, the authors in \cite{aich2022protecting}  presented a model based on the FL and blockchain, 
which is 
used to predict COVID-19 symptoms, how it spreads, and speed up using the medical data in research and treatment. 
In addition, the combination of FL and blockchain could be useful for real time environment and for organizations that do not want share sensitive data with third parties because of privacy concerns. After analyzing the combination of blockchain and FL solution, the authors noted that the proposed solution securely protects the data access and would help to build a robust model.

Similarly to~\cite{samuel2022iomt}, Lakhan et al.~\cite{lakhan2022federated} proposed a privacy-preserving FL for IoMT system. It is a mathematical model called FL-BETS, which is a FL-based privacy enhancing and malware detection-enable blockchain IoMT system
for 
different healthcare
workloads. The aim of this study is to preserve privacy and fraud of data in the local fog nodes and remote clouds network  with minimum energy consumption and delay. The performance evaluation of the FL-BETS framework, compared to other existing machine learning and blockchain methods in malware  analysis shows the best performance  in fraud analysis, data validation, energy and delay constraints for healthcare applications. Also, the model decreases energy consumption by 41\%
and delay by 28\%.


Similar to~\cite{samuel2022iomt} and~\cite{lakhan2022federated}, in~\cite{singh2022framework}, the authors proposed the model by integrating Blockchain and FL-enabled approaches to provide a secure architecture for privacy preservation in smart healthcare systems.  In this model, blockchain-based IoT cloud apps enhance security and privacy by combining FL and blockchain technologies. The proposed model has provided secure data sharing for the IoHT environment with privacy preservation. Organizations can use federated-based blockchain cloud architecture without collaborating sensitive and private healthcare system data in the cloud.

Kumar et al. \cite{kumar2021blockchain} developed an FL blockchain-based approach to train the global model for detection 
of 
COVID-19 patients based on Computed tomography (CT) slices while preserving the privacy of patients' private data and the organization, similarly to~\cite{samuel2022iomt}. The proposed model evaluated real-life COVID-19 patients' data\footnote{https://paperswithcode.com/dataset/cc-19} that were collected from various hospitals with different types of CT scanners and publicly available to the research community. The results showed that the blockchain-based FL smartly detects COVID-19 patients using computed tomography (CT) scans among various hospitals while preserving sensitive data privacy. 
Table~\ref{blockchain} summarizes 
the blockchain methods applied for FL in smart healthcare.

\section{Key challenges for future research}
\label{KEY}

While PETs in FL have many advantages and have been growing rapidly in recent years, some challenges cannot be ignored. Existing frameworks are still at an early stage and need improving methods to enhance data privacy. 

\subsubsection{Computation cost}
One of the main challenges of FL is represented by privacy-enhancing to prevent data leakage. FL needs multiple iterations to achieve the final global model. Therefore, the number of training iterations has a direct impact on increasing the cost of the training model. As shown in 
\cite{rachakonda2022privacy}, multi-party computation is a way to protect data privacy in FL. Performing experiments with a different number of workers does not impact the computation cost, however, increasing the number of training rounds significantly 
boosts 
the computation cost. Therefore, the trade-off between privacy risk and computation time has been a promising topic for researchers.

\subsubsection{Privacy and security}

In Section~\ref{subsec:blockchain}, some studies show that integration of the blockchain method and FL is a way to enhance privacy in IoHT. However, there is an open issue that may lead to privacy leakage. In the FL, only the central server has information about the sources of the local model updates, and the addresses of the clients are private. However, addresses in blockchain are public, and using blockchain in FL gives the ability to other clients to communicate with each other and obtain the training model based on the public information from the blockchain. Therefore, the risk of data leakage among clients cannot be ignored.

\subsubsection{Linkage attacks} 
The k-anonymity technique is a way to preserve the anonymity of individuals. The key idea is how to modify the
attributes of the dataset in a way that each instance has at
least $k$-1 other entities with identical quasi-identifiers. Therefore, an identifiable record would link to multiple records in the anonymous dataset. However, k-anonymity cannot avoid privacy leakage against linkage attacks where a sensitive attribute is shared among a group of individuals with the same quasi-identifier.

\section{Conclusions}
\label{con}

This survey has reviewed nineteen representative works that apply federated learning (FL) in the Internet of Healthcare Things (IoHT) domain in terms of privacy aspects, including attacks and privacy-enhancing technologies (PETs). The datasets used by these works have also been summarized, which are helpful for researchers aiming to reproduce these works. 
Some open research issues on the topic still exist, such as the trade-off between privacy risk and computational time, the risk of data leakage among colluding clients, and the sharing of sensitive attributes.
Along with the current research efforts, we encourage more insights into the problems of this area and more efforts in addressing the open research issues identified in this paper.


\section*{Acknowledgments}

This research is part of the INCT of the Future Internet for Smart Cities funded by CNPq proc. 465446/2014-0, Coordena\c{c}\~ao de Aperfei\c{c}oamento de Pessoal de N\'ivel Superior – Brasil (CAPES) – Finance Code 001, FAPESP proc. 14/50937-1 and proc. 15/24485-9.

\balance
\bibliographystyle{IEEEtran}
\bibliography{mybibfile}
\end{document}